\documentclass[aps,prl,reprint,superscriptaddress]{revtex4-2}
\usepackage{graphicx}
\usepackage{subfigure}
\usepackage{amsmath}
\usepackage{amssymb}
\begin{document}

% Use the \preprint command to place your local institutional report
% number in the upper righthand corner of the title page in preprint mode.
% Multiple \preprint commands are allowed.
% Use the 'preprintnumbers' class option to override journal defaults
% to display numbers if necessary
%\preprint{}

\title{Anomalous In-plane Magnetic Anisotropy in Strain-mediated Converse Magnetoelectric Coupling}

% repeat the \author .. \affiliation  etc. as needed
% \email, \thanks, \homepage, \altaffiliation all apply to the current
% author. Explanatory text should go in the []'s, actual e-mail
% address or url should go in the {}'s for \email and \homepage.
% Please use the appropriate macro foreach each type of information

% \affiliation command applies to all authors since the last
% \affiliation command. The \affiliation command should follow the
% other information
% \affiliation can be followed by \email, \homepage, \thanks as well.

\author{Zefeng Cai}
\email[Electronic address: ]{czf17@mails.tsinghua.edu.cn}
%\homepage[]{Your web page}
%\thanks{}
%\altaffiliation{}
\affiliation{School of Materials Science and Engineering, Tsinghua University, Beijing 100084, China}

% \collaboration{MUSO Collaboration}%\noaffiliation
\author{Ben Xu}
%  \homepage{http://www.Second.institution.edu/~Charlie.Author}
\affiliation{School of Materials Science and Engineering, Tsinghua University, Beijing 100084, China}

\date{September 7, 2020}

\begin{abstract}
% insert abstract here
Magnetic axis rotation (MAR) in ferromagnetic (FM) layers is crucial for strain-mediated converse magnetoelectric coupling. Employing the density functional theory (DFT), we computationally study the magnetic anisotropy of selected deformed FM materials such as body-centered iron. The results show that the short axis is more energy-favorable at high in-plane strain difference than previously predicted phenomenologically. This anomalous trend and the complex energy behaviors at different strain conditions explain why spin-lattice dynamics (SLD) simulation does not produce in-plane MAR and imply couplings between different energy terms together with high order coefficient contributions.
\end{abstract}

% insert suggested keywords - APS authors don't need to do this
%\keywords{}

%\maketitle must follow title, authors, abstract, and keywords
\maketitle
% body of paper here - Use proper section commands
% References should be done using the \cite, \ref, and \label commands
\section{Background and Purpose}
% Put \label in argument of \section for cross-referencing
%\section{\label{}}
Magnetoelectric (ME) coupling and ME heterostructures designed based on it are of growing interest these days. With advantages such as non-volatility, low energy cost, and excellent controllability, electrically tunable magnetic devices provided with this kind of heterostructures attract both academia and industry. ME heterostructures appear as a dielectric (DE) layer (usually piezoelectric (PE) or ferroelectric (FE)) contacting a ferromagnetic (FM) layer\cite{RN99,RN101}. In such a structure, how electric factors propagate through DE layers and affect the magnetization in FM layers (i.e., the mechanisms of converse ME coupling) is necessary for understanding existing experimental data and designing new material of promising functions. 

Recent studies have focused on strain-mediated converse ME coupling, one of the mechanisms of converse ME coupling, which exists in most heterostructures consisting of PE layers. A specific strain state in the PE layer is induced by a given electric field in this mechanism and can propagate to the FE layer. The strain breaks down the symmetry of the layer, which changes the magnetic anisotropy of FE layers. Experimental results have implied the existence and controllability of this mechanism. J. Z. Cui et al. use patterned electrodes on a PE substrate to generate a localized strain to control the magnetic anisotropy in a thin Ni island on the substrate, which is measured based on the magneto-optical Kerr effect (MOKE)\cite{RN100}. The distance from electrodes and islands, which contributes to the magnitude of the strain, can dramatically affect the magnetic anisotropy. A. W. Rushforth et al. study the magnetic anisotropy of a Mn-doped GaAs Hall rod bonded to a PZT piezotransducer\cite{RN107}. By detecting unconventional crystalline components of the anisotropic magnetoresistance (AMR), they demonstrate the voltage, which leads to different strain patterns, control the direction of the easy axis.

Strain-mediated converse ME coupling receives attention because it has the potential to realize tricky functions resorting to various engineering routes. For example, it can be utilized to design electric-field-driven $180^\circ$ magnetization switching, which counts in magnetoelectric random access memories (MeRAM)\cite{RN101}. R. C. Peng et al., using a finite element analysis (FEA) and phase-field model, computationally depict a $180^\circ$ in-plane magnetization switching on a CFB/PZT heterostructure by a time-dependent pulse voltage\cite{RN106}. The appearance and disappearance of precession play an essential role during switching. J. J. Wang et al., employing similar phase field simulation, demonstrate a $180^\circ$ in-plane magnetization switching on Ni/PMN-PT heterostructure utilizing four-fold symmetric anisotropy of a nanomagnet through two deterministic $90^\circ$ switchings\cite{RN110}.

However, how strain microscopically interacts with FM layers and induces magnetization switching, especially in-plane switching, remains a serious question. This gap limits the prediction for the property of complicated ME heterostructures and unintuitive design of complex ME material. Several theoretical approaches and models are employed to explain the behavior of strain-mediated ME heterostructures. J. M. Hu et al.\cite{RN102} develop a phenomenological model for EAR (easy axis reorientation) of this mechanism calculating the total free energy change, which is divided into the magnetocrystalline anisotropy energy, the shape anisotropy, and the magnetoelastic energy. They apply the model to some specific PE/FM-layered heterostructures and classify such EARs into out-of-plane and in-plane ones. The easily-calculated numerical model developed by J. Z. Cui et al. is similar to Hu’s model for the in-plane case\cite{RN100}. Cui’s results based on the model remarkably agree with their FEA results. However, they assume these separated energy terms maintain their forms in the total energy expression. The chances are that some terms of energy are coupled and cannot be regarded as independent variables. Besides, we cannot confirm that a specific divide covers all components of total energy. 

Additionally, the completeness of the model depends on the experimental results of specific materials. It hinders the application of the model to heterostructures using materials that we have not been able to fabricate or whose parameters we have difficulty in measuring. To eliminate the dependence on empirical parameters in phenomenological models, A. W. Rushforth et al. combine Luttinger Hamiltonian using fixed band parameters with the kinetic exchange model using fixed exchange parameters, trying to find the dependence of the total energy density on the magnetization in-plane angle at a given strain state\cite{RN107}. This approach, without major dependence on empirical parameters, still requires an experimental value of $K_u$ to fix one adjustable parameter: the shear strain along the reference axis. Besides, the energy density distribution calculated only indicates the thermodynamic equilibrium state, lacking information on the dynamic process. 

Phase-field model, whose primary process is to solve the Landau-Lifshitz-Gilbert equation, is utilized by Peng and Wang\cite{RN106,RN110}. On the one hand, it can explicitly exhibit the dynamic switching process, through which we can trace the trajectory of every small elementary part and roughly predict the switching time-scale. On the other hand, it remains the drawback of the phenomenological model, because the term of total free energy F still exists in the effective magnetic field. Additionally, the microscopic subunits in the simulation conducted are just “phantom” spins, which cannot sense any real atomic force field. Scientists apply specific parameters of different materials on the model only by modifying the form of total free energy, which means it is impossible to capture the fundamental physical process based on quantum mechanics behind the phenomena.

Lacking atomistic simulation technique to process systems with spins, scientists in preceding research have difficulty in taking both lattice potential and exchange integral into consideration. P. W. Ma et al.\cite{RN104} develop a spin-lattice-electron evolving model by adding the Heisenberg and Landau terms to classical molecular dynamics (MD) Hamiltonian. They release an open spin-lattice dynamics code based on this model. They employ it and successfully depict the propagation of compressive waves through iron and some other spin-lattice systems. Based on this technique, C. P. Chui and Y. Zhou\cite{RN96} successfully illustrate the external field effect on BCC iron and its reinforcement on both long-range and short-range magnetic order at high temperature. The appearance of new simulating tools inspires and accelerates our research.

In this paper, we regard the strain-mediated converse ME coupling as a kind of spin-lattice coupling and research it by more microscopic simulation and calculation. Based on the spin-lattice molecular dynamics (SLD) approach, we simulate the MAR response of body-centered iron under different strain conditions. Employing the density functional theory (DFT), we computationally study the magnetic anisotropy of body-centered (BCC) iron and face-centered (FCC) nickel. Comparing the results with those from the previous phenomenological model, we find anomalous in-plane magnetic anisotropy in selected FM materials during strain-mediated converse ME coupling.

\section{Results and Discussion}
In exhibition and analysis below, we always assume the normal strain difference (NSD), which means the difference between the strain along X-axis and that along Y-axis, positive due to the symmetry of the systems we study. That means we can name the X-axis “the long axis” and the Y-axis “the short axis” for convenience. 

\begin{figure}[h]
\centering
\includegraphics[width=0.48\textwidth]{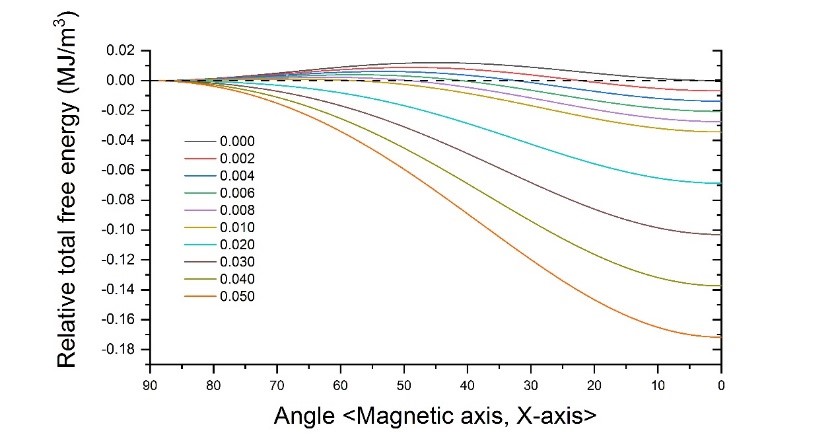}
\caption{300 K in-plane magnetic anisotropy with difference normal strain difference (NSD) calculated by J. M. Hu et al.’s phenomenological model\cite{RN102}. The horizontal axis shows the values of the included angles between the magnetic axis and the X-axis given in the degree units. The vertical axis shows the total free energy change relative to the situation where the magnetic axis lays along the long short Y-axis given in $\text{MJ/m}^3$ units. Different colors of the lines represent different NSD given in the units of one.\label{fig1}}
\end{figure}

FIG.~\ref{fig1} shows different included angles between the magnetic axis and the long axis (the X-axis) graphed against the total free energy change at 300 K, which we calculated using J. M. Hu et al.\cite{RN102}’s magnetization anisotropy model (Eq.~(\ref{eq6})).

It implies if we apply a particular value of strain to a BCC iron (001) epitaxial film, the magnetic moment that is initially along the short axis can rotate to the long axis direction. We employ the magnetocrystalline coefficients of BCC iron measured by P. Escudier\cite{RN131}, the magnetostriction coefficients of BCC iron at 300 K measured by G. M. Williams et al.\cite{RN129}, and the stiffness coefficients measured by A. E. Lord et al\cite{RN128}. When the included angle is smaller than $45^\circ$, the closer the magnetic axis to the long axis of the system, the lower the total free energy. An inspiring fact is that if NSD is larger than the critical value, the free energy barrier between the situation where the magnetic axis parallels to the long axis and the situation where the magnetic axis parallels to the short axis vanishes. This critical value is approximately 0.01, according to Fig.~\ref{fig1}. Considering the theorem that the systems are always seeking for the low free energy state, it implies if we apply the strain to a BCC iron (001) epitaxial film, satisfying the condition that NSD is larger than 0.01, the magnetic moment that is initially along the Y-axis can rotate to the X-axis direction without hindrance. We name that in-plane magnetic axis rotation (MAR).

%\begin{figure}
%\includegraphics[width=0.5\textwidth]{exchange.jpg}		
%\caption{The diagram of different initial spin configurations we employed in is fixed at 0.93874 per atom. The values of spin are dimensionless.\label{fig2}}
%\end{figure}

\begin{figure}[h]
\includegraphics[width=0.45\textwidth]{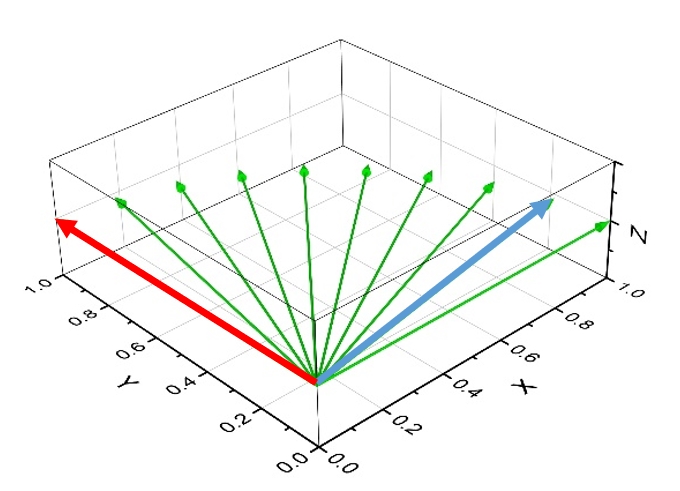}
\caption{The diagram of different initial spin configurations we employed in spin-lattice dynamics simulation. The X-axis is the long axis, while the Y-axis is the short axis as is discussed before. The cube represents the (001) film of BCC iron in the simulation box. Green unit vectors represent the different initial total spin directions in the simulation box. The average longitudinal magnitude of initial spin is fixed at 0.93874 per atom. The values of spin are normalized to be dimensionless.\label{fig2} }
\end{figure}

We conduct the spin-lattice dynamics (SLD) simulation to produce the phenomena of in-plane MAR. Fig.~\ref{fig2} shows the initial spin configurations of our system. The simulation has the NSD of 0.02. If J. M. Hu et al.’s model is correct\cite{RN102}, the situation where the initial magnetic axis parallels to the Y-axis (the red vector in Fig.~\ref{fig2}) is enough to exhibit the rotation of the magnetic axis. We set other nine uniform transition states in case the free energy barrier still exists, and we can study the local free energy distribution by analyzing the simulating results of these transition configurations. 

\begin{figure}[h]
\includegraphics[width=0.45\textwidth]{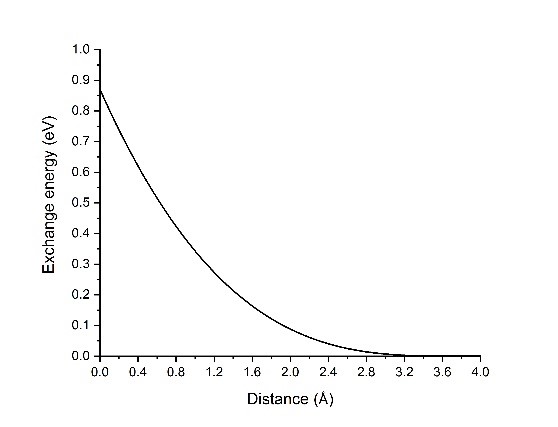}
\caption{Exchange function $J\left(\mathbf{R}\right)$ in Heisenberg exchange term fitted by P. W. Ma et al.\cite{RN104} in their spin-lattice Hamiltonian. The horizontal axis shows the increasing distance between two atomic spins given in the angstrom units. The vertical axis shows the exchange energy of these two spins given in the eV units.\label{fig3} }
\end{figure}

We can qualitatively understand the existence of in-plane MAR under large NSD in a classical way. Assuming the longitudinal magnitude of atomic spin is approximately a constant before and after strain, which is valid if the strain is not big enough to change the density of state dramatically, the magnitude of Heisenberg exchange term determines the energy favorability according to Eq.~(\ref{eq1}) \cite{RN104},
\begin{equation}
H_{spin} = -\frac{1}{2}\sum_{i,j}J_{ij}(\mathbf{R})\mathbf{S}_i \cdot \mathbf{S}_j\;.\label{eq1}
\end{equation}
The larger the exchange term, the more stable the spin configuration. The exchange function $J(\mathbf{R})$ in the exchange term always decreases with increasing distance R between two atomic spins, which is implied by the curve fitting by P. W. Ma et al.’s \cite{RN104} (see Fig.~\ref{fig3}). If every single atomic spin is approximately parallel to each other, which is the essential property of ferromagnetic materials, the longer the axis the total spin lies along, the closer every two atomic spins. Therefore, the configuration where the magnetic axis parallels to the long axis is more energy-favorable than the configuration where the magnetic axis parallels to the short axis.

\begin{figure*}
\subfigure[$90^\circ$]{%
\includegraphics[width=0.49\textwidth]{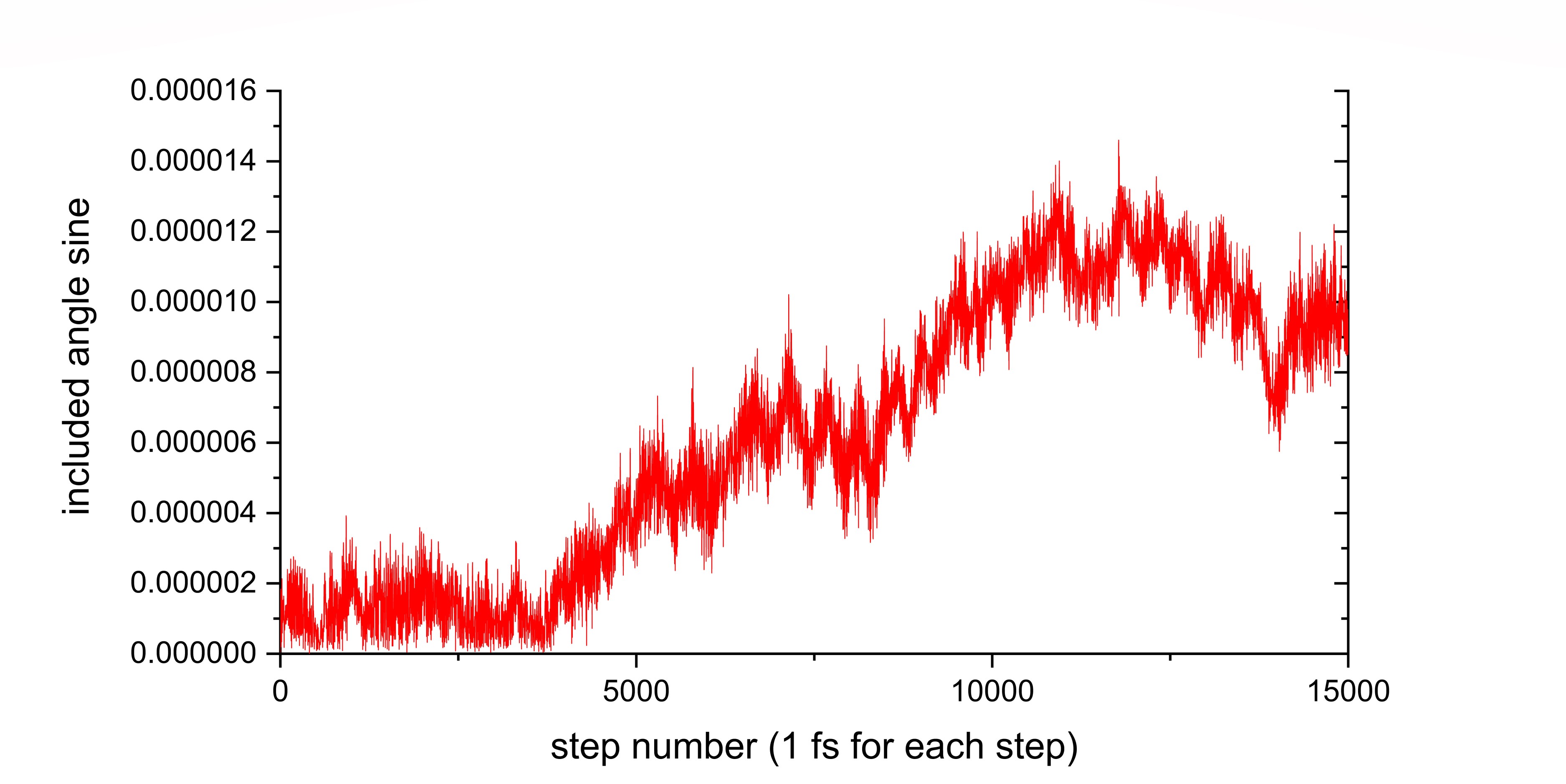}\label{fig4:a}}
\subfigure[$10^\circ$]{%
\includegraphics[width=0.49\textwidth]{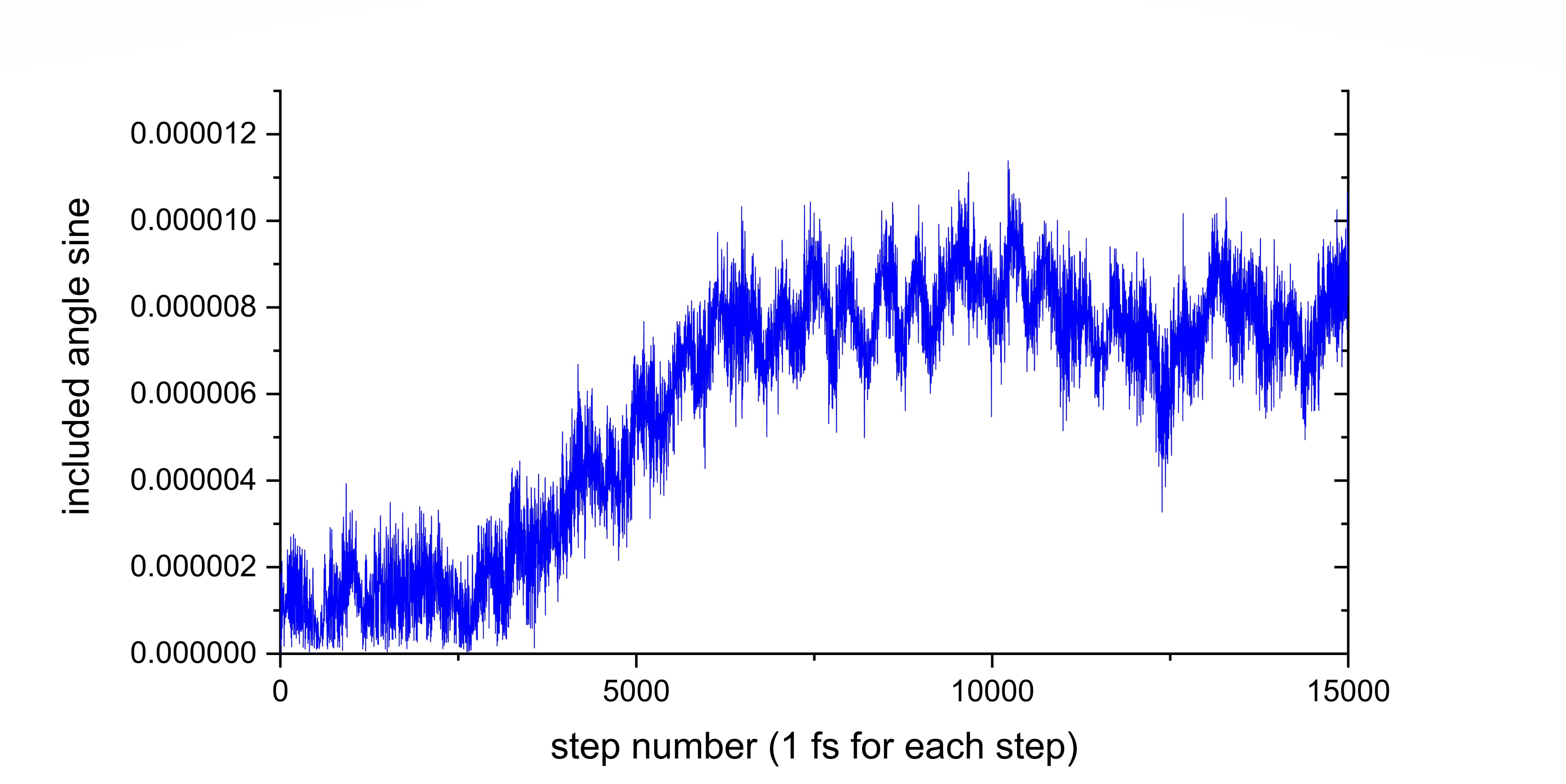}\label{fig4:b}}
\caption{\ref{fig4:a} The sine value of the included angle between the initial total spin and the total spin at the given steps. The initial angle between the total spin and the X-axis is $90^\circ$, which is marked as the red vector in FIG.~\ref{fig2}. The step size of the spin-lattice dynamics we conducted is one fs. The curve only exhibits the thermal vibration of spins. \ref{fig4:b} The sine value of the included angle between the initial total spin and the total spin at the given steps. The initial angle between the total spin and the X-axis is $10^\circ$, which is marked as the blue vector in FIG.~\ref{fig2}. }
\label{fig4}
\end{figure*}

We calculated the sine of angles between the initial total spin and the total spin at given simulating steps. The expected in-plane MAR does not appear. Fig.~\ref{fig4:a} shows the sine deviation of the system with the initial total spin of a $90^\circ$ angle with the long axis (the red vector in Fig.~\ref{fig2}) and Fig.~\ref{fig4:b} shows that with the initial total spin of a $10^\circ$ angle with the long axis (the blue vector in Fig.~\ref{fig2}). The in-plane MAR, however, does not occur in the simulations with all kinds of initial configurations. Even though Fig.~\ref{fig4:a} and Fig.~\ref{fig4:b} only exhibits the deviation within 15 ps, the deviation curve shows out-of-order fluctuation in a more extended range. The maximum of sine deviation is less than $1.5 \times 10^{-5}$. The atomic spins are right at their original sites and only oscillate as the result of the finite temperature (the NVT ensemble is employed). The results of SLD simulation indicates that the energy distribution in between the long axis and the short axis is different from that predicted by the phenomenological theory. Because all coefficients exist in Eq.~(\ref{eq6}) are measured in the mechanical equilibrium state and treating the summation of all components of free energy as the total free energy only makes sense in equilibrium state, in order to eliminate the impact from non-equilibrium, we set the normal strain along the Z-axis from 0 to -0.0117, which is calculated by the mechanical equilibrium condition (Eq.~(\ref{eq2},\ref{eq3}))\cite{RN102},

\begin{equation}
\sigma_{ij} = \frac{\partial{\left(\Delta{F_{me}} + \Delta{F_{el}}\right)}}{\partial{\varepsilon}_{ij}}\;,\label{eq2}
\end{equation}
\begin{equation}
\varepsilon_{33} = -\frac{\left[B_1\left( m_3^2-\frac{1}{3}\right)+c_{12}\left(\varepsilon_{11} + \varepsilon_{22}\right)\right]}{c_{11}}\;.\label{eq3}
\end{equation}

However, the in-plane MAR still does not appear. Two possible explanation is that the rotation has a much larger energy barrier or the energy change is too flat or locally fluctuating to generate an MD-observable rotation. The fact that in-plane MAR does not happen in any transition configuration in the time scale of 1 ns at most implies that the free energy curve with positive NSD is very flat or locally fluctuating, namely the latter explanation.

\begin{figure*}
\subfigure[Phenomenological Model]{%
\includegraphics[width=0.49\textwidth]{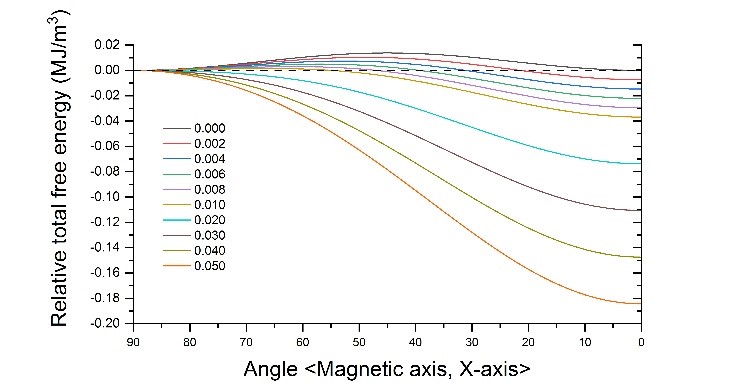}\label{fig5:a}}
\subfigure[DFT]{%
\includegraphics[width=0.49\textwidth]{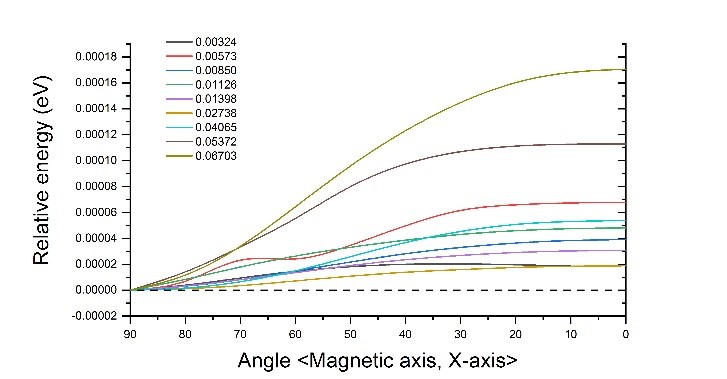}\label{fig5:b}}
\caption{\ref{fig5:a} 4.2 K in-plane MAE of BCC iron bulk with difference NSD calculated by J. M. Hu et al.’s phenomenological model, which is the same as Fig.~\ref{fig1}. \ref{fig5:b} 0 K in-plane MAE of BCC iron bulk with difference NSD calculated by DFT with SOC. }
\label{fig5}
\end{figure*}

\begin{figure*}
\subfigure[Phenomenological Model]{%
\includegraphics[width=0.49\textwidth]{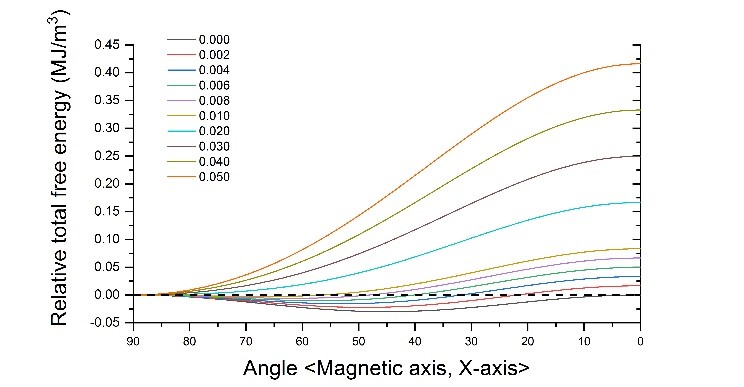}\label{fig6:a}}
\subfigure[DFT]{%
\includegraphics[width=0.49\textwidth]{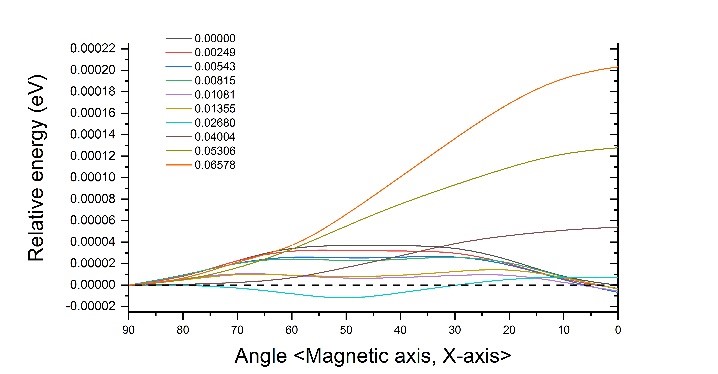}\label{fig6:b}}
\caption{\ref{fig6:a} 4.2 K MAE of FCC nickel bulk with difference NSD calculated by J. M. Hu et al.’s phenomenological model. \ref{fig6:b} 0 K in-plane MAE of FCC nickel bulk calculated by DFT with SOC. }
\label{fig6}
\end{figure*}
Employing density functional theory (DFT), we calculate the in-plane magnetic anisotropy energy (MAE) of BCC iron bulk (Fig.~\ref{fig5:b}) and FCC nickel bulk (Fig.~\ref{fig6:b}) to understand the precise magnetic anisotropy curve. Because the energy shifts of the bands induced by spin-orbit interaction are considered as the origin of the magnetic anisotropy in ferromagnetic materials13, we consider the effect of spin-orbit coupling. The local magnetic moment in a given volume is calculated by Eq.~(\ref{eq4}),
\begin{equation}
\overrightarrow{M_I} = \int_{\Omega_I} \overrightarrow{m}(\mathbf{r})F_I(\left|\mathbf{r}\right|)\,d\mathbf{r}\;,\label{eq4}
\end{equation}
where $F_I$ is a function of norm one inside $\Omega_I$, which smoothly goes to zero towards the boundary of $\Omega_I$. The system is also optimized to the mechanical equilibrium by ionic relaxation. We also calculate the phenomenological magnetic anisotropy of BCC iron (Fig.~\ref{fig5:a}) bulk and FCC nickel (Fig.~\ref{fig6:a}) bulk for comparison. Even if the phenomenological magnetic anisotropy model is proposed for epitaxial film, the term of shape anisotropy is totally canceled in the in-plane case and the total free energy expression for films is still correct for bulks. 

Additionally, because 4.2 K is near 0 K, the entropy of the system can be ignored, and the curve calculated by DFT can be treated as the total free energy change, which is the same as the curve generated by the phenomenological model. Therefore, Fig.~\ref{fig5:a} and Fig.~\ref{fig5:b} can be compared, as well as Fig.~\ref{fig6:a} and Fig.~\ref{fig6:b}. We assume the MAE varies continuously by the NSD, which is usually valid when studying the properties of three-dimensional bulk materials.

The DFT-SOC results of BCC iron show that when the NSD is larger than 0.003, the closer the magnetic axis to the long axis, the higher the total free energy change, which is in stark contrast with the phenomenological results. The 4.2 K phenomenological MAE shows the similar trend to the 300 K situation (i.e., the closer the magnetic axis to the long axis, the lower the total free energy when the included angle is smaller than 45o.) The MAE calculated by DFT indicates that the in-plane MAR never happens in BCC iron with any value of NSD. The MAE of the phenomenological model can agree with that of DFT only when the NSD smaller than 0.003. 

The DFT-SOC results of FCC nickel exhibit a complicated trend with the increasing NSD. When the NSD is smaller than 0.0815, the MAE has an energy barrier between the long axis and the short axis. When the NSD is larger than about 0.01 and smaller than about 0.03, the MAE exhibits a decreasing trend. When the MSD is larger than 0.04, the closer the magnetic axis to the long axis, the higher the total free energy change. Since the FCC nickel has a negative magnetostriction coefficient (i.e., shrinking when a magnetic field is applied) in contrast with the BCC iron, the 4.2 K phenomenological MAE shows like the inverted reflection of that of the BCC iron. Since either the rotation barrier or negative energy difference between the short axis direction and the long axis direction always exists according to Fig.~\ref{fig6:b}, we can also speculate that the in-plane MAR never happens in FCC nickel with any value of NSD. The MAE of the phenomenological model agrees with that of DFT only when the NSD larger than 0.04. However, the particular values of the total free energy change differ a lot. The total free energy change calculated by DFT is 0.00047 eV per cell, while that calculated by the phenomenological model is around 0.00014 eV per cell after unit conversions. 

The DFT MAE of iron and nickel shows like the combination of the phenomenological MAE of both of them. The curve with NSD of 0.0573 in Fig.~\ref{fig5:b} exhibits the trend similar to the stage where the NSD is smaller than 0.015 in Fig.~\ref{fig6:a}. The curves with NSD smaller than 0.00815 in Fig.~\ref{fig6:b} exhibit the trend similar to the stage where NSD is smaller than 0.01 in Fig.~\ref{fig5:a} It suggests the complex coupling between different free energy terms together with high order coefficient contribution plays a significant role in in-plane MAE. The reason the Monte Carlo simulation conducted by T. V. Dung et al.\cite{RN130} and the phase field simulation conducted by J. J. Wang et al.\cite{RN110} exhibit the MAR of the BCC iron and the FCC nickel respectively is that they only consider the first order of corresponding coefficients, and the results are semi-empirical.

\begin{figure*}
\subfigure[Iron]{%
\includegraphics[width=0.4\textwidth]{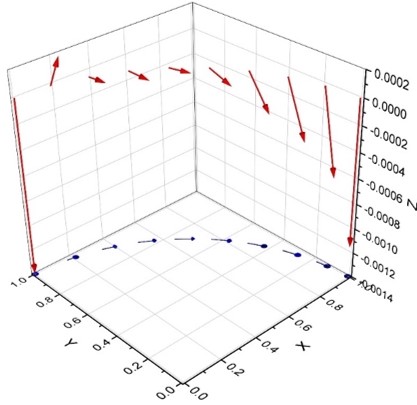}\label{fig7:a}}
\qquad
\subfigure[Nickel]{%
\includegraphics[width=0.4\textwidth]{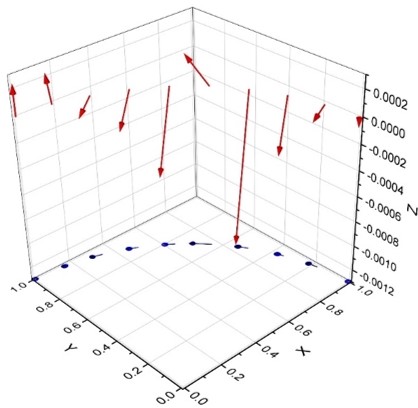}\label{fig7:b}}
\caption{The three-dimensional diagram of local spin change during SCF iteration. The X-axis is the long axis, while the Y-axis is the short axis. Lattice vector along the Z-axis remains constant. The red vectors point to the final direction vectors of local spin from the initial direction vectors. The blue vectors are the projection of conversion vectors on the XY-plane. Discrete vectors represent the local spin change of different initial spin configurations. \ref{fig7:a} The local spin change of BCC iron bulk. \ref{fig7:b} The local spin change of FCC nickel bulk. }
\label{fig7}
\end{figure*}

Additionally, the local spin change during the self-consistent field (SCF) iteration of DFT remains a question (Fig.~\ref{fig7}). The spin moves from the in-plane direction to the out-of-plane direction dramatically in contrast to the phenomenological model, which declares that the out-of-plane rotation is much more difficult to happen in the BCC iron. It implies that the three-dimensional MAE should be considered when predicting the occurrence of magnetic axis rotation. Future research can focus on three-dimensional MAE with different NSD. The XY-plane projections of the change vectors of iron (Fig.~\ref{fig7:a}) and nickel (Fig.~\ref{fig7:b}) have converse orientations while the decreasing of energy has the same orientation. Whether it is a numerical phantom or some physical manifestations also remains a future orientation. Because VASP has poor support for thermodynamic correction, the free energy at given finite temperature rather than 0 K energy obtained from VASP is of small significance. The future work is also supposed to be creating a precise method to determine the free energy correction. The experiments measuring the iron and nickel MAE with different NSD is still lacking. Future experiments results are expected to be a persuasive material for strain-mediated converse ME coupling mechanism.

\section{Conclusion}

The first-principle magnetic anisotropy has a complicated trend for BCC iron and FCC nickel. It is dramatically different from that predicted by J. M. Hu et al.’s phenomenological model without high order contribution and coupling between different energy terms. The real magnetic anisotropy must have a flatter or locally fluctuating distribution according to spin-lattice dynamics.

\appendix
\section{Detail of phenomenological model and calculation}
We study the in-plane easy axis reorientation (EAR) of single-domain BCC iron and single-domain FCC nickel under different strain patterns by a phenomenological model. Two kind of films are grown epitaxially along (001) orientation by assumption. We adopt the phenomenological anisotropy model built by J. M. Hu et al.\cite{RN102} to calculate the critical value of strain, with which the in-plane spontaneous EAR has no energy barrier. The total free energy change of a single-domain FM film can be described as the sum of magnetocrystalline anisotropy energy change, $\Delta{F_{mc}}$ magnetostatic energy change $\Delta{F_{ms}}$ and magnetoelastic energy change $\Delta{F_{me}}$ (Eq.~(\ref{eq5})),
\begin{equation}
\Delta{F_{tot}} = \Delta{F_{mc}} + \Delta{F_{ms}} + \Delta{F_{me}}\;.\label{eq5}
\end{equation}
Given that the system we study is a single-domain film, magnetostatic energy change above is only contributed by shape anisotropy change. Additionally, we are only interested in the in-plane case, where the direction cosine of the easy axis with respect to the out-of-plane principal axis is constantly zero. Then we plug the expression of each term into the formula above and obtain the total free energy contribution under this circumstance,
\begin{equation}
\Delta{F_{tot}} = K_1 m_1^2 \left(1-m_1^2\right)-\frac{3}{2}\lambda_{100}\,\Delta\varepsilon\left(c_{11}-c_{12}\right)m_1^2\;,\label{eq6}
\end{equation}
where $m_1$ is the direction cosine of the easy axis with respect to the X-axis, $\Delta{\varepsilon}$ is the difference between the value of normal strain along X-axis and that along Y-axis (i.e., NSD). $K\,,\lambda\,,c$ are specific components of bulk anisotropy constants, magnetostrictive constants, and elastic stiffness constants respectively. We employ the magnetocrystalline coefficients of FCC nickel at 4.2 K measured by P. Escudier\cite{RN131}, the magnetostriction coefficients measured by J. J. M. Franse\cite{RN127}, and the stiffness coefficients calculated by finite-differences phonon calculation in VASP, while the coefficients of BCC iron have the same source of Fig.~\ref{fig1}. We plug in these parameters and calculate the critical value of NSD, where the coefficient in front of the quadratic term is zero (i.e., the total free energy change has no maximum larger than 0), and regard it as the critical value of strain without energy barrier for EAR.

\section{Detail of spin-lattice dynamics simulation}
We study the behavior of the BCC iron bulk with normal strain by a spin-lattice dynamics (SLD) program SPILADY written by P. W. Ma et al.8 The simulation is based on the equations of motion derived from a special Hamiltonian (Eq.~(\ref{eq7})),
\begin{widetext}
\begin{equation}
%\begin{split}
H = \left[\sum_i \frac{\mathbf{p}_i^2}{2m_i} + U\left(\{\mathbf{R}_i\}\right)\right]-\left[\frac{1}{2}\sum_{i,j}j_{ij}\left(\mathbf{R}_{ij}\right)\mathbf{e}_i \cdot \mathbf{e}_j + H_L \right]\;,\label{eq7}
%\end{split}
\end{equation}
% put long equation here
\end{widetext}

where the first term is the classical molecular dynamics (MD) Hamiltonian. The second term is the spin Hamiltonian, including Heisenberg Hamiltonian and Landau Hamiltonian. The whole equations of motion based on this Hamiltonian treat both the spins and the atoms, as well as the interaction between them. A film-like simulation box consisting of $200\times200\times15$ BCC conventional unit cells 46.6nm$\times$46.6nm$\times$4.25nm  is built. The DD05 iron potential function (one of EAM potential functions), which serves as the $U\left(\{\mathbf{R}_i\}\right)$ term in the Hamiltonian above, and the program-embedded Heisenberg exchange function and Landau coefficients are employed in our simulation. Our system is treated as a canonical ensemble (NVT) at 300 K, which is realized by Langevin thermostats based on the fluctuation-dissipation theorem. The initial total spin direction is along Y-axis. Different initial patterns of strain with NSD ranging from 0.005 to 0.05 are applied at the first simulating step. The normal strain along Y-axis is always set as zero for convenience. We monitor the trajectories of each spin in the box through the default output of SPILADY program. It is worth noting that all patterns of strain we applied eliminate the energy barrier if the EAR occurs, according to the critical condition we calculated above.

\section{Detail of \emph{ab initio} approach}
We study the spin structures of BCC iron and FCC nickel by using a plane wave pseudopotential approach to the density functional theory (DFT), which is implemented in the Vienna \emph{Ab initio} Simulation Package (VASP). We employ the generalized gradient approximation (GGA) with Perdew-Burke-Ernzerhof (PBE) exchange-correlation functional. The projector augmented wave (PAW) pseudopotentials we used in the calculation includes eight valence electrons for Fe ($4s^23d^6$) and ten valence electrons for Ni ($4s^23d^8$). The wavefunctions are expanded in a plane-wave basis with a cutoff energy of 600 eV (for the ionic relaxation) and 450 eV (for the static calculation). A $17 \times 17 \times 17$ Monkhorst-Pack k-point sampling centering on Gamma point is used for one conventional unit cell of iron. We build a series of re-optimized unit cells whose initial NSDs ranging from 0 to 0.05. The spin-orbit coupling is taken into consideration. The magnetic anisotropy energy (MAE) of these structures is obtained by continuously changing the quantization axis for noncollinear spins (i.e., the SAXIS label in INCAR). We calculate the single-point energy of each configuration. The direction of the quantization axis is distributed uniformly on the space between the +x vector and the +y vector on the XY-plane. Each self-consistent electronic calculation is converged to $10^{-6}$ eV (for the ionic relaxation) and $10^{-6}$ eV (for the static calculation) and the tolerance force is set to 0.005 eV/\AA (for the ionic relaxation). The occupation numbers and the spin-orbit coupling matrix elements are obtained from the output file PROCAR.

% If you have acknowledgments, this puts in the proper section head.
%\begin{acknowledgments}
% put your acknowledgments here.
%\end{acknowledgments}

% Create the reference section using BibTeX:
\bibliographystyle{apsrev4-2}
\bibliography{spin_lattice}

\end{document}